\documentclass[epjCONF,columns]{svjour} 
\usepackage{graphics}
\usepackage[varg]{txfonts} 
\usepackage[latin1]{inputenc}
\usepackage{color}
\usepackage{cancel}

\session-title{LHCP 2013}
\begin{document}
\title{Distinguishing between lepton number violating scalars at the LHC}
\author{Francisco del Aguila\inst{1}\fnmsep\thanks{\email{faguila@ugr.es}} 
\and Mikael Chala\inst{1} \fnmsep\thanks{\email{miki@ugr.es}} 
\and Arcadi Santamaria\inst{2} \fnmsep\thanks{\email{arcadi.santamaria@uv.es}} 
\and Jose Wudka\inst{3}\fnmsep\thanks{\email{jose.wudka@ucr.edu}} }
\institute{CAFPE and Departamento de F{\'\i}sica Te{\'o}rica y del Cosmos, Universidad
de Granada, E\textendash{}18071 Granada, Spain 
\and Departament de F{\'\i}sica Te\`orica, Universitat de Val{\`e}ncia and IFIC,
Universitat de Val{\`e}ncia-CSIC, Dr. Moliner 50, E-46100 Burjassot (Val{\`e}ncia),
Spain \and Department of Physics and Astronomy, University of California, Riverside CA 92521-0413, USA}
\abstract{
Scalars with lepton number violating interactions decaying into lepton pairs,
as those mediating the see-saw of type II, always include
doubly-charged components. If these are observed at the LHC,
their electro-weak quantum numbers can be determined 
through their leptonic decays in pair and single production.
} 
\maketitle
\section{Introduction}
\label{intro}
LHC data, and especially the discovery of the Higgs boson, 
\cite{Aad:2012tfa,Chatrchyan:2012ufa} 
have confirmed the validity of the minimal Standard 
Model (SM) below few hundreds of GeV, 
except for the very tiny neutrino masses. 
No other signal of new physics (NP) has been observed 
up to now, in general pushing the gap to the next NP scale 
beyond the TeV.  
Then, one may also wonder if the remaining SM predictions 
are fulfilled. In particular, at what extent the accidental 
symmetries of the SM, like lepton number (LN), are also 
exact, and if eventually the observation of their violation is 
within the LHC reach. 

Although neutrino masses are so small, the mere 
observation of that they are non-vanishing requires 
the extension of the SM. 
Thus, if neutrinos are Dirac particles, 
we have to add their right-handed (RH) counterparts, 
leaving LN conserved; 
while if they are Majorana, new particles mediating LN violating (LNV) 
processes must be added. 
Which, if light enough, may be observable at the LHC 
\cite{Keung:1983uu} 
{\footnote{In general, they also generate extra 
contributions to neutrinoless double $\beta$ decay, 
besides the standard one proportional to the 
electron neutrino Majorana mass; which are conveniently 
parametrized by the corresponding low energy effective operators 
\cite{Schechter:1981bd,Babu:2001ex,Choi:2002bb,Engel:2003yr,de Gouvea:2007xp,delAguila:2012nu,delAguila:2013zba}.}}. 
The simplest realization of this scenario results from 
the addition of heavy Majorana neutrinos, giving to the 
SM neutrinos a mass through the {\it see-saw} mechanism, 
named of type I 
\cite{Minkowski:1977sc,Yanagida:1979as,GellMann:1980vs,Glashow:1979nm,Mohapatra:1979ia}. 
The observation of these extra neutral leptons may be 
problematic at the LHC because in the absence of other 
interactions they are produced through their mixing with 
the SM leptons 
\cite{Han:2006ip,del Aguila:2007em,Atre:2009rg}, 
which is bounded to be small \cite{delAguila:2008pw}. 
Moreover, the corresponding SM backgrounds are also 
rather large at hadron colliders.

As a matter of fact, the less known sector within the SM 
is the scalar one; and if there is NP at the LHC reach, 
it is likely to be related with the Higgs sector. 
The most popular SM extension by the addition of new 
scalars violating LN is provided by the see-saw of 
type II, which gives neutrinos a Majorana mass through the 
exchange of an electro-weak scalar triplet of hypercharge 1, 
$(\Delta^{++}, \Delta^+, \Delta^0)$, 
coupling to lepton doublet as well as to gauge boson pairs, 
and hence with no well-defined LN 
\cite{Marshak:1980yc,Schechter:1980gr,Cheng:1980qt}. 
Stringent limits on the mass of $\Delta^{\pm\pm}$ in the 200-400 GeV range 
have been already set by CMS 
\cite{Chatrchyan:2012ya} and ATLAS \cite{ATLAS:2012hi}, 
assuming that this doubly-charged scalar mainly decays into 
two same-sign leptons, and allowing for a variety of particular cases 
with fixed branching ratios into different lepton flavours 
{\footnote{In definite models as in the see-saw of type II, the 
Yukawa couplings giving neutrinos a mass are the same 
mediating the like-charge di-leptonic scalar decay, and they 
are then constrained \cite{Hektor:2007uu,delAguila:2008cj}, 
but this is not so in general.}}. 
The mass bounds on $m_{\Delta^{\pm\pm}}$, however, 
depend strongly on the choice for these branching ratios. 
Moreover, in both analyses the decay rate into 
$W^\pm W^\pm$ is assumed to be negligible, as are 
also the possible LNV signals. 

Here we review the generalization of the type II  see-saw scenario, 
where we consider the addition of scalar multiplets with a priori arbitrary isospin 
$T = 0, 1/2, 1, \cdots$, and hypercharge $Y$, and no well-defined LN 
\cite{delAguila:2013yaa} 
{\footnote{Models with isosinglets of hypercharge 2 have been 
proposed, for instance, in 
\cite{Zee:1985id,Babu:1988ki,delAguila:2011gr,Gustafsson:2012vj}; 
whereas the phenomenological implications of isodoublets 
with hypercharge 3/2 are studied in 
\cite{Gunion:1996pq,Rentala:2011mr,Aoki:2011yk,Yagyu:2013kva}.}}.
In general, branching ratios into same-sign lepton and gauge 
boson pairs are also assumed to be sizable; and therefore, LNV 
eventually observable. 
It must be emphasized, however, that in the see-saw of 
type II, and in more elaborated models, both branching ratios 
are only similar in a small region of parameter space. 
Generally the doubly-charged scalars decay almost exclusively 
into same-sign lepton pairs; or alternatively into gauge bosons, 
though in this case the signal cannot be extracted from the background 
for the range of masses under consideration 
\cite{Kanemura:2013vxa}. 

Our analysis has also a different purpose than the one guiding 
the discovery of a doubly-charged scalar $H^{\pm\pm}$ decaying 
into two like-charge leptons, and which has been carried out to derive 
present experimental limits on the corresponding cross-sections. 
Present bounds are obtained assuming a mass for $H^{\pm\pm}$ 
and some fixed branching ratios to same-sign lepton pairs of 
a given flavor composition. Then, the comparison of the 
predicted number of events including SM backgrounds with the 
observed one provides the corresponding limit, as no departure 
from the SM prediction is observed. 
In contrast, we assume that the doubly-charged scalar has been 
already observed to resonate in the same-sign di-lepton channel 
$l^\pm_1l^\pm_2$, with $l_{1,2} = e,\mu$, in which case we describe a process 
for measuring its branching ratios to $ll, l\tau, \tau\tau$ and $WW$, as well as its total 
pair production cross-section. 

\section{Scalar production and decay}
\label{sec:1}

Any doubly-charged scalar $H^{\pm\pm}$ showing a resonant behaviour  
in the invariant mass distribution of same-sign di-leptons 
must couple to one of the two SM bilinears, 
$\overline{L^c_{L}} L_{L}$ and $\overline{l^c_{R}} l_{R}$, 
{\footnote{$L^c_L = (\nu^c_L, l^c_L)$ is the SM lepton doublet 
with charge-conjugated fields, 
$\psi^c_L = (\psi_L)^c = C\overline{\psi_L}^T$, $\psi^c_R = (\psi_R)^c = C\overline{\psi_R}^T$.}} 
with LN = 2.  
{\footnote{The other combination $\overline{L^c_{L}} l_{R}$ requires 
a $\gamma^\mu$ insertion because of the fermions' chirality, 
and hence the presence of a covariant derivative $D_\mu$ to make the operator 
Lorentz invariant. The corresponding operators are equivalent to the ones 
constructed with the other two bilinears through the use of the 
equations of motion.}} 
There being no restriction on the hypercharge $Y$ and on the isospin $T$ 
of the electro-weak multiplet it belongs to, except that one of its components 
must be doubly-charged: $|2-Y| \leq T$. 
Since we can write gauge invariant operators with this multiplet 
and any of the two LN = 2 lepton bilinears by including enough SM 
(charge-conjugated) Higgs doublets $\phi = (\phi^+, \phi^0)$  
($\tilde{\phi} = i\sigma_2 \phi^*$, with $\sigma_2$ the second 
Pauli matrix). 
These in turn give rise to the effective coupling (decay) of the 
doubly-charged scalar (in)to the two same-sign charged leptons, 
after electro-weak symmetry breaking when the Higgs acquires a 
vacuum expectation value (VEV) $<\phi^0> = v/\sqrt2 \approx 174$ GeV.

In contrast with its leptonic decay, which only depends on 
the size of the effective coupling of the corresponding operator 
and is then a priori a free parameter, 
the $H^{\pm\pm}$ production is fixed by gauge invariance and 
hence does depend on the electro-weak multiplet it belongs to, 
that is, on $T$ and $T^{H^{++}}_3 = 2 - Y$.

\subsection{Doubly-charged scalar pair and single production}
\label{sec:1.1}

At a hadron collider $H^{\pm\pm}$ are mainly pair produced 
through the $s$-channel 
exchange of photons and $Z$ bosons, being their electro-weak 
gauge couplings obtained from the corresponding kinetic term 
$(D_\mu H)^\dagger (D^\mu H)$. They can be also be produced singly 
through the s-channel exchange of $W$ bosons if $T \neq 0$ . 
Both cross-sections are of electro-weak size but vary with the multiplet $H^{\pm\pm}$ 
belongs to \cite{delAguila:2013yaa}. For instance, they grow 
with $T$ for multiplets with components of charge at most 
$Q = T_3 + Y = 2$ 
(that is, with $T^{H^{++}}_3 = T$). 
They can be also produced by vector boson fusion, 
which can be considered to be a next-order 
correction and will be neglected here 
{\footnote{Single doubly-charged production through 
vector boson fusion can be eventually at the LHC reach 
for rather bizarre models evading the stringent electro-weak 
constraints on the VEV of the neutral scalar partner of the 
doubly-charged scalar boson \cite{Chiang:2012dk,Englert:2013wga}. 
In this case the doubly-charged scalar only decays into W pairs, and 
then it does not resonate in the di-leptonic channel.}}. 
Although the vector boson fusion contribution grows with the 
$H^{\pm\pm}$ mass, 
it is smaller than the one mediated by the s-channel electro-weak 
gauge boson exchange by almost an order of magnitude for 
multiplets satisfying $T^{H^{++}}_3 = T$ and 
doubly-charged scalar masses below the TeV; 
details will be provided in a forthcoming publication. 
In any case, the corresponding events have two extra 
jets with relatively large pseudo-rapidity which provide an extra 
handle to separate them.

Our purpose here is to review how to determine the 
doubly-charged scalar pair production cross-section, and thus 
measure $T^{H^{++}}_3$, using only four lepton 
final states. We concentrate for simplicity on the case 
$T^{H^{++}}_3 = T$, 
{\footnote{Models also including scalars with larger electric charges 
can produce even more striking signals 
\cite{Babu:2009aq,Bambhaniya:2013yca,Franceschini:2013aha}. 
In any case, we assume that mass splittings within multiplets 
due to mixing with other scalars are small, and can be 
neglected \cite{Akeroyd:2011ir,Aoki:2011pz}.}} 
thus restricting $T = 0, 1/2, 1, 3/2, 2$. 

\section{Sampling}
\label{sec:2}

Large hadron colliders have become precision machines 
due to the excellent performance of the detectors and the high 
energy and luminosity reached. Thus, the understanding of the SM 
backgrounds seems to be well beyond initial expectations, 
and the LHC potential, for instance, for observing new resonances 
in leptonic channels is foreseen to be well beyond the TeV region in next 
LHC runs, although their properties can be only established 
for lower masses. 

Here we review the example discussed in Ref. \cite{delAguila:2013yaa}, 
where we assume a heavy scalar with a mass $m_{H^{\pm\pm}} = 500$ GeV, 
belonging to a weak singlet ($T=0$), doublet ($T=1/2$), triplet ($T=1$), 
quadruplet ($T=3/2$) or quintuplet ($T=2$) of hypercharge $Y = 2 - T$, 
and a LHC center of mass energy $\sqrt s = 14$ TeV. 
Doubly-charged scalar pairs are then generated using MADGRAPH5 
\cite{Alwall:2011uj}, after implementing the corresponding vertices 
for the different multiplet assignments (see Eqs. (3-4) in Ref. \cite{delAguila:2013yaa}) 
in Feynrules \cite{Christensen:2008py}, 
and the CTEQ6L1 parton distribution functions. 
Backgrounds are evaluated with ALPGENV2.13 \cite{Mangano:2002ea}, \break
while parton radiation and fragmentation are simulated with 
PYTHIAV6 \cite{Sjostrand:2006za} and the detector with DELPHESV1.9 \break 
\cite{Ovyn:2009tx}. 
In the analyses, leptons have a minimum transverse 
momentum  $p_T^l > 15$ GeV and a maximum pseudo-rapidity $|\eta_l| < 2.5$.

\subsection{Pair production cross-section determination}
\label{sec:2.1}

We are interested in measuring the total cross-section $\sigma$ 
in order to determine the isospin of the multiplet the doubly-charged 
scalar belongs to. However, this cannot be done using a single 
decay mode because the corresponding number of signal events also 
depends on the $H^{\pm\pm}$ branching ratio into this mode, 
which is in principle unknown. 
Thus, we must determine the number of events for each channel 
and sum them appropriately taking into account the different 
efficiencies to estimate the total cross section. 
What can be done using only events with four charged leptons, properly sampled, 
and Eq. (7) in Ref. \cite{delAguila:2013yaa} is to measure 
\begin{equation}
\sigma = \left(\sigma_{ll ll} + \frac{1}{2}\sum_{a \neq ll} \sigma_{ll a} \right) ^2 / \sigma_{ll ll}
\, , 
\label{eq:sigma}
\end{equation}
where $\sigma_{ab} = (2-\delta_{ab})\sigma z_a z_b$ is the 
$H^{\pm\pm}$ pair production cross-section into $ab$, with 
$z_a \equiv Br( H \rightarrow a)$ the $H$ branching ratio into $a$ 
and $\sum_{a = ll, l\tau, \tau\tau, WW} z_a = 1$. 

In practice we proceed as follows. We select events with four 
charged leptons of the first two families with zero total electric charge, 
plus possibly missing transverse momentum, and 
require that at least one same-sign pair 
$l_1l_2$ reconstructs the scalar mass within $\pm 40$ GeV. 
(See Table \ref{tab:1} for a summary of the main cuts.)
\begin{table}
\caption{Main applied cuts.}
\label{tab:1}      
\begin{tabular}{lccc}
\hline\noalign{\smallskip}
Variable & $l^\pm l^\pm l^\mp l^\mp$ & $l^\pm l^\pm l^\mp\tau^\mp$ &  
$l^\pm l^\pm\tau^\pm\tau^\pm (W^\pm W^\pm)$\\\hline
\# of leptons & 4 & 4 & 4\\
$|m_{l_1^\pm l_2^\pm}-m_{H^{\pm\pm}}|$ & $\le 40$ GeV & $\le 40$ GeV & $\le 40$ GeV\\
$|m_{l_3^\pm l_4^\pm}-m_{H^{\pm\pm}}|$ & $\le 40$ GeV & $\ge 40$ GeV & $\ge 40$ GeV\\
$\cancel{p}_T$ & not applied & $\ge 50$ GeV & $\ge 50$ GeV\\
$x_3$ & not applied & $\ge 0.8$ & $\le 0.7$\\
\noalign{\smallskip}\hline
\end{tabular}
\end{table}
These events are then separated into three disjoint sets depending 
on the category associated to the other like-sign pair: $ll, l\tau$ and 
$\tau\tau + WW$. 
We denote by $ll$ those events resulting from the decay of the second 
doubly-charged scalar into two leptons of the first two families, and hence, 
those with the second like-sign lepton pair $l_3l_4$ also reconstructing the 
scalar mass within $\pm 40$ GeV. 
For the remaining events we assume that $l_3$ and $l_4$ 
are products of semileptonic tau decays, and distribute 
the transverse missing momentum of the event between both tau 
leptons with the requirement that their momenta align along the 
momentum of the corresponding product charged-lepton momentum it decays to: 
$xp_\tau^\mu = p_l^\mu$, with $0 < x < 1$. 
Then, we identify the event with the second scalar decaying 
into a lepton of the first two families and a tau lepton, $l\tau$, as those 
fulfilling that the fraction momentum of the most energetic 
of the two leptons not reconstructing the scalar mass, 
which we name $l_3$, ($x_3$) is larger than 0.8. 
The other events are classified as resulting from the decay 
of the second scalar into two tau leptons or two gauge bosons, 
$\tau\tau + WW$.

\begin{figure}
\resizebox{1.16\columnwidth}{!}{%
\includegraphics{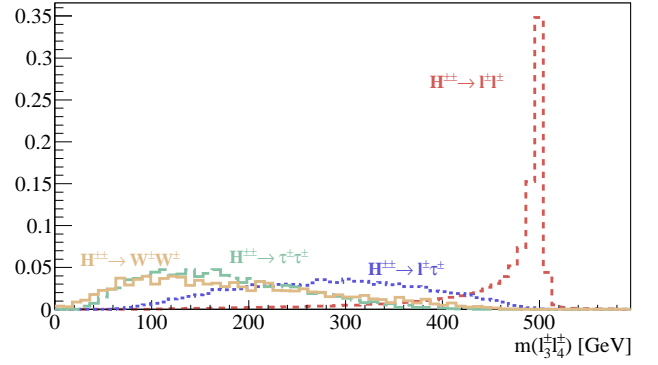} }
\caption{Invariant mass of the two same-sign leptons 
that provides the poorest reconstruction of  the 
$H^{\pm\pm}$ mass in doubly-charged scalar 
pair production for different decay modes.}
\label{fig:1}
\end{figure}

\begin{figure}
\hspace{-0.21cm}\resizebox{1.16\columnwidth}{!}{%
\includegraphics{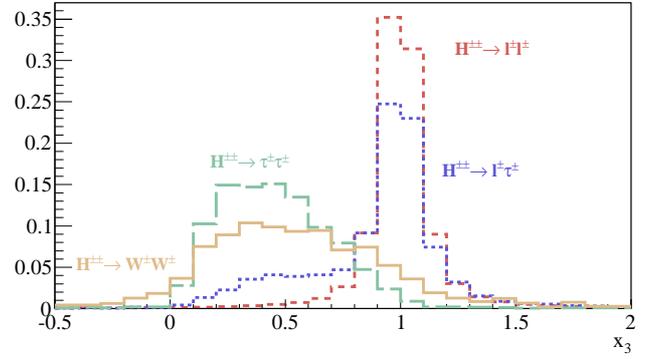} }
\caption{Momentum fraction, $x_3$, of the most energetic lepton of the 
two which worst reconstruct the $H^{\pm\pm}$ mass in doubly-charged scalar 
pair production for different decay modes.}
\label{fig:2}
\end{figure}

\begin{figure}
\resizebox{1.16\columnwidth}{!}{%
\includegraphics{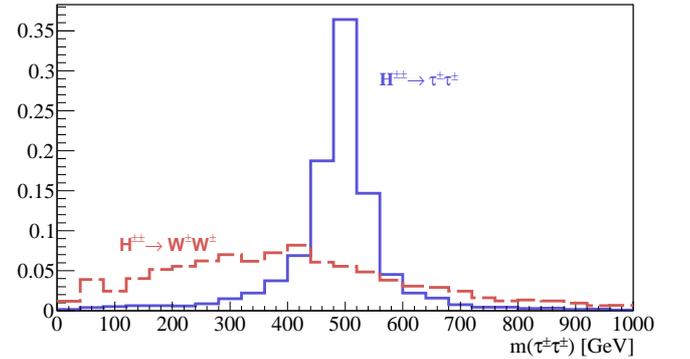} }
\caption{Invariant mass distribution of the two reconstructed taus 
for both $l^{\pm}l^{\pm}W^{\mp}W^{\mp}$ and 
$l^{\pm}l^{\pm}\tau^{\mp}\tau^{\mp}$ decay channels in 
doubly-charged scalar pair production.}
\label{fig:3}
\end{figure}

Figures \ref{fig:1}, \ref{fig:2} and \ref{fig:3} prove the usefulness of the 
procedure. Indeed, 
Figure \ref{fig:1} shows the like-sign di-lepton invariant 
mass for the lepton pair  which provides the poorest reconstruction of the 
doubly-charged scalar mass and for its four decay modes: 
$H^{\pm\pm} \rightarrow l_3^\pm l_4^\pm$, 
$H^{\pm\pm} \rightarrow l_3^\pm \tau^\pm \rightarrow l_3^\pm l_4^\pm {\cancel p}_T$, 
$H^{\pm\pm}  \rightarrow \tau^\pm \tau^\pm \rightarrow l_3^\pm l_4^\pm {\cancel p}_T$, 
and 
$H^{\pm\pm}  \rightarrow W^\pm W^\pm \rightarrow l_3^\pm l_4^\pm {\cancel p}_T$. 
The separation of the first sample is obtained simply requiring  
$m_{l_3^\pm l_4^\pm}$ to be in between $m_{H^{\pm\pm}} \pm 40$ GeV. 
In Figure \ref{fig:2} we plot the momentum fraction assigned to the 
most energetic lepton of the second like-sign pair for the former four types 
of scalar decays. As can be observed, $x_3 \geq 0.8$ provides a rather clean 
separation of the first two decay modes from the last two. 
The different distributions reflect the fact that the leading lepton tends to 
be the one which is not a decay product of a $\tau$ or a $W$, and thus 
with $x_3 \sim 1$.
In order to further reduce the contamination we also require $x_3 \leq 0.7$ 
for $\tau\tau + WW$ events. 
Finally, Figure \ref{fig:3} makes it clear that separating 
$\tau\tau$ and $WW$ decays is rather inefficient. 
However, what further justifies not doing this at this stage, 
for evaluating the total pair production cross-section, is 
that the efficiency $\epsilon$ for evaluating the cross-sections of both types 
of processes is rather similar, allowing to treat both together consistently.  

\begin{figure}
\resizebox{1.065\columnwidth}{!}{%
{\includegraphics{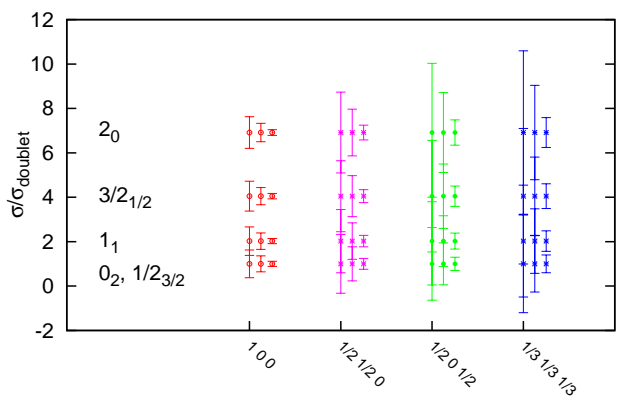}} }
\caption{Uncertainty on the determination of 
$\sigma(pp\rightarrow H^{++}H^{--})$ for different 
$H^{\pm\pm}$ branching ratio assumptions at three 
different luminosities: 100, 300 and 3000 fb$^{-1}$. 
The labels on the left stand for the doubly-charged 
scalar $T_Y$ quantum numbers.}
\label{fig:4}
\end{figure}

Actually, in order to measure the total pair production cross-section 
using Eq. (\ref{eq:sigma}), we have to estimate the efficiency 
for each process. 
From our Monte Carlo simulations we have calculated 
the efficiency for each subsample including the corresponding
branching ratios, obtaining: 
$\epsilon_{llll} = 0.6, \epsilon_{lll\tau} = 0.09$ and 
$\epsilon_{ll\tau\tau} = \epsilon_{llWW} = 0.02$, respectively.
Then, counting the number of events of each of the three 
subsets and dividing by the corresponding efficiency, 
we can measure the doubly-charged scalar pair production 
cross-section, once the integrated luminosity $\cal L$ is known. 
In Figure \ref{fig:4} we plot the statistical error for such 
a determination for several $H^{\pm\pm}$ branching 
ratio assumptions and the five multiplet assignments $T_Y$ discussed above. 
The cross-sections are correctly reproduced because they 
are an input, but what matters is with which precision can we measure 
them in order to distinguish between different multiplet assignments.  
We draw statistical errors including the effect of SM backgrounds for three 
different integrated luminosities: 100, 300 and 3000 fb$^{-1}$. 
(For example, we find $\sim 50$ background events for ${\cal L} = 300$ fb$^{-1}$ 
with four charged leptons adding to zero total electric charge and 
a same-sign pair reconstructing the $H^{\pm\pm}$ mass within 40 GeV.) 
We fix four definite sets of doubly-charged scalar branching ratios:  
$(z_{ll}, z_{l\tau}, z_{\tau \tau}+z_{WW}) =$ $(1, 0, 0)$, 
$(1/2, 1/2, 0)$, $(1/2, 0, 1/2)$, $(1/3, 1/3, 1/3)$, being higher the precision 
with $z_{ll} > z_{l\tau} > z_{\tau \tau}+z_{WW}$ . 
This means worsening when worse is our ability to reconstruct the 
$H^{\pm\pm}$ mass. 

\begin{figure}
\resizebox{1.17\columnwidth}{!}{%
\includegraphics{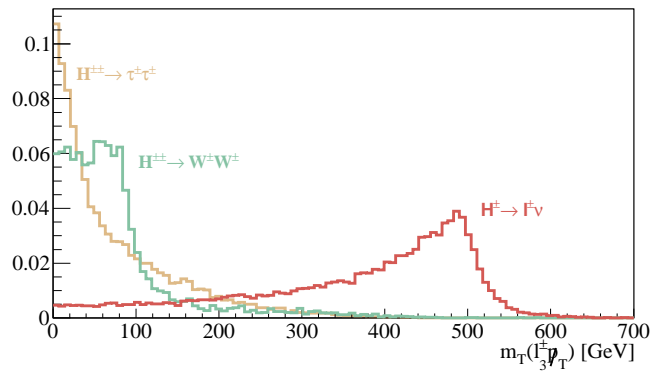} }
\caption{Transverse mass distribution of the opposite-sign lepton and the 
missing tranverse momentum for several signals.}
\label{fig:5}
\end{figure}

As can be observed in Figure \ref{fig:4}, there is no apparent difference 
between the singlet and doublet cases because their 
neutral cross-sections are rather similar, but this is not 
so for the charged ones as there is no charged-boson exchange 
graph in the singlet case \cite{delAguila:2013yaa}. 
Thus, the associated production of doubly-charged scalars can 
be used to discriminate between these two multiplets. 
If there is a significant excess of events compatible with 
$H^{\pm\pm}H^{\mp}$, the singlet hypothesis will be 
automatically ruled out. However, to establish whether this 
is the case must be carefully assessed because 
the observation of only three leptons in the final state 
does not uniquely characterize this (charged-current) 
process. For instance, pair produced doubly-charged 
scalars decaying into 
$l^\pm l^\pm l^\mp \tau^\mp, l^\pm l^\pm \tau^\mp \tau^\mp$ 
or $l^\pm l^\pm W^\mp W^\mp$ can also produce 
only three leptons if one $\tau$ or $W$ decays hadronically.  
Therefore, we need to rely on extra variables in order to forbid 
these (neutral-current) contributions. 
Two discriminators appear to be most convenient: 
the missing transverse momentum, ${\cancel p}_T$, 
which is larger in single $H^{\pm\pm}$ production than in 
$l^\pm l^\pm l^\mp \tau^\mp$ events with 
$\tau^\mp$ decaying hadronically, which allows to separate 
both contributions; and the transverse invariant mass of 
the opposite-sign lepton and the missing transverse momentun, 
$m_T(l^\pm_3 {\cancel p}_T)$, 
which peaks near the $H^{\pm}$ mass in the charged-current 
process, as shown in Figure \ref{fig:5}. 
It is apparent from this figure that a cut on $m_T > 250$ GeV 
suppresses the neutral $\tau\tau$ and $WW$ contributions quite efficiently. 

Although it is neither efficient nor necessary to separate $\tau\tau$ from 
$WW$ events in order to determine the total $H^{\pm\pm}$ pair production 
cross-section in Eq. (\ref{eq:sigma}), one must attempt to do it in order to establish or not 
the violation of LN at the LHC. This separation makes use of the 
invariant  mass distribution in Figure \ref{fig:3}: $WW$ events are 
defined as those outside a wide enough interval around the $H^{\pm\pm}$ 
mass. Reconstructed $\tau\tau$ events near the doubly-charged scalar 
mass are interpreted as genuine $\tau\tau$ decays. For the events left, 
and interpreted as $WW$ decays, one must also check that the assignment 
is consistent with such a $H^{\pm\pm}$ decay. 
One can also check the consistency with the excess of events 
resulting from the semi-leptonic decay of the $WW$ pair, but 
this has other backgrounds, too, and will be discussed elsewhere.

\section*{Acknowledgements}

This work has been supported in part by the Ministry of Economy and 
Competitiveness (MINECO), under the grant numbers FPA2006-05294, 
FPA2010-17915 and FPA2011-23897, by the Junta de Andaluc{\'\i}a 
grants FQM 101 and FQM 6552, by the 
``Generalitat Valenciana'' grant PROMETEO/2009/128, and by the 
U.S. Department of Energy grant No.~DE-FG03-94ER40837. 
M.C. is supported by the MINECO under the FPU program.

  
\end{document}